# Gene Expression Patterns of *CsZCD* and Apocarotenoid Accumulation during Saffron Stigma Development

**Zohreh Shams**

Department of cell molecular and physiology, Alzahra University, Tehran, Iran.

## ABSTRACT

*Crocus sativus* L., otherwise known as saffron, is a highly prized plant due to its unique triploid capability and elongated stigmas, contributing to its status as the costly spice globally. The color and taste properties of saffron are linked to carotenoid elements including cis- and trans-crocin, picrocrocin, and safranal. In the research carried out, we dedicated our attention to the gene *CsZCD*, an important player in the formation of apocarotenoids. Through the application of real-time polymerase chain reaction to RNA purified from saffron stigmas at various growth phases, it was determined that the peak expression of the *CsZCD* gene coincided with the red stage, which is associated with the highest concentration of apocarotenoids. The data showed a 2.69-fold enhancement in *CsZCD* gene expression during the red phase, whereas a 0.90-fold and 0.69-fold reduction was noted at the stages characterized by orange and yellow hues, respectively. A noteworthy observation was that *CsZCD*'s expression was three times that of the *CsTUB* gene. Additionally, relative to *CsTUB*, *CsLYC* displayed 0.7-fold and 0.3-times expression. Our investigation provides insight into the governance of *CsZCD* during stigma maturation and its possible influence on the fluctuation in apocarotenoid content. These discoveries carry significance for the industrial production of saffron spice and underscore the importance of additional studies on pivotal genes participating in the synthesis of apocarotenoids.

## INTRODUCTION

Saffron (Crocus sativus) is a fragrant herb believed to have descended from C. *cartwrightianus* through natural evolution. Its cultivation for thousands of years has produced a valuable crop known for its triploid potency and long stigmas, making it the world's most expensive spice (Freeman et al., 1999; Livak et al., 2001; Bouvier et al., 2001). Saffron is best found in carotenoid components such as cis- and trans-crocin, picrocrosine and saffron, which give it its characteristic flavoring and coloring properties (Rubio et al., 2008; Mir et al., 2012b).

Clonal selection has emerged as an important strategy for improving saffron cultivation, as other methods such as mutagenesis and chromosome doubling have had limited success. Identification and propagation of the best clones with improved stain properties and improved carotenoid biosynthetic capacity are important for the improvement of saffron crops (Sanchez et al., 2013b), These strategies should be deployed more in these situations that global warming and drought stress put plants survival in jeopardy (Mir et al., 2012b: Jamshidi et al., 2022).

The intricate process involved in the development of saffron's distinctive color and flavor components is a result of the bio-oxidative cleavage of zeaxanthin. This process is orchestrated by critical enzymes encoded by genes such as PSY, LYC, CCD, BCH, and ZCD. The primary color compound, crocin, and the aromatic substance, safranal, are hypothesized to result from the bio-oxidative division of zeaxanthin through a 7,8 (7',8') cleavage reaction (Pfander and Schurtenberger, 1982). The apocarotenoid biosynthetic pathway encompasses numerous enzymes which catalyze reactions and are encoded by key genes such as LYC, PSY, BCH, ZCD, and, CCD. Lycopene β-cyclase (LYC) facilitates the cyclization of lycopene, leading to the formation of β-carotene with two rings. The β-carotene hydroxylation in the MVA pathway is facilitated by β-carotenoid hydroxylase, encoded by the BCH gene, resulting in the production of zeaxanthin (Castillo et al., 2005). The production of color and aroma in saffron results from the bio-oxidative cleavage of zeaxanthin (Rubio-

# Gene Expression Patterns of CsZCD and Apocarotenoid Accumulation during Saffron Stigma Development

Moraga et al., 2009; Gomez-Gomez et al., 2010) at the 7,8 (7',8') sites by the zeaxanthin cleavage dioxygenase (*CsZCD*), producing crocetin dialdehyde and picrocrocin. The terminal phase in C. sativus stigmas involves the glucosylation of the cleavage products of zeaxanthin by the glucosyltransferase 2 enzyme, which is encoded by the *CsUGT2* gene within the chromoplast of stigmas. These products are then stored in the central vacuole of the fully matured stigmas (Bouvier et al., 2003). This understanding of the biosynthetic procedure can shed light on the maturation process of saffron (Baghalian et al., 2010; Ahrazem et al., 2010).

Soil electrical conductivity (EC) plays an important role in saffron cultivation, as lower EC levels are associated with higher yields (Mirbakhsh and Hosseinzadeh, 2013; Ahrazem et al., 2015; Abdulhabip et al., 2017). Examining the effect of soil electrical conductivity on bulb and flower properties, as well as the effect of saffron tissue culture on gene expression and physical activity, could improve saffron production to meet global demand (Baba et al., 2017: Bagheri et al., 2017; Mirbakhsh et al., 2023).

Given the many medicinal and commercial uses of saffron, the manufacturing process may not be adequate. For this reason, biotechnological methods such as tissue culture are being investigated for dissemination. Culture induced stigma-like structure (SLS) has become an important step in the production of coarse saffron, and studies on its effects have helped refine the process (Frusciante et al., 2014; Jain et al., 2016; Gomez-Gomez et al. 2017). Saffron's fascinating history and complex chemical composition, cultivation concept and propagation methods make it the world's most important spice. The aim of this study is to patterns of the expression of essential genes that are involved in apocarotenoid synthesis and evaluate accumulation of apocarotenoid by comparing *CsZCD* gene expression during three stages of stigma development (Figure 1).

## MATERIAL AND METHODS

### Plant preparation
A total of twenty-five different saffron clones collected from the saffron germplasm of Mashhad province, Iran was used in this study. The clones were carefully preserved at the Central Temperate Pasteur Institute in Tehran, Iran. To preserve their integrity, freshly harvested heads were quickly removed in liquid nitrogen and stored at -80°C for RNA isolation for RNA isolation.

### Extraction procedure
### RNA extraction
Stigmas in their three developmental phases (yellow, orange, and scarlet) were frozen and subsequently ground into a fine powder using a chilled, sterilized mortar and pestle. This finely ground material was then subjected to the extraction of total RNA employing a kit designed for RNA isolation, provided by Roche Applied Sciences, Penzberg, Germany. This process was conducted in strict accordance with the guidelines provided by the manufacturer in the kit.

### cDNA preparation
Each specimen was processed with 5 µg of the whole RNA serving as a template, followed by the execution of first-strand cDNA synthesis. This procedure utilized an 18-bp oligo dT primer and a first-strand cDNA synthesis kit supplied by Roche Applied Science, located in Penzberg, Germany, all while strictly adhering to the instructions provided by the manufacturer. The synthesized cDNA was subsequently preserved at a temperature of -20°C, reserved for future utilization in gene expression research.

### Assessing the quality of the strands
The quality of the RNA that was extracted, evaluated by determining the absorbance at 260 and 280 nm by a Nano-Drop spectrophotometer. RNA showcasing an optical density (OD) ratio of 260/280 between 1.2 and 1.5 was selected for cDNA synthesis. The first strand cDNA was synthesized using 5 µg of the total RNA template and an 18 bp oligo-dT primer in conjunction with a cDNA synthesis kit supplied by Roche Applied Sciences, Penzberg, Germany, as per the supplier's guidelines. The resulting cDNA was preserved at -20°C for later utilization. The Beer-Lambert law, which provides a correlation between absorbance and concentration, was employed to ascertain the RNA concentration. An A260 reading of 1.0 corresponds to approximately 40 µg/ml of RNA, and absorbance at 260 nm was utilized to gauge the RNA concentration in solution. The purity of the RNA preparation was assessed based on the absorbance ratio of pure RNA between 260 and 280 nm, where an A260/A280 ratio is 2.1. The Nano-Drop® ND-1000 UV-Vis Spectrophotometer, which eliminates the necessity for cuvettes and capillaries, thereby reducing the required sample volume, was utilized for efficient analysis of small samples. To eliminate carotenoids, saffron stigmas were subjected to treatment with methanol followed by Tris-HCl (pH 7.5; containing 1 M NaCl). The precipitate was collected via centrifugation, ground again in acetone, and centrifuged. This procedure was repeated until the pellet was devoid of color. The supernatants were then amalgamated, evaporated, and the resulting dried residue was stored at -80°C for future use.

### Real Time-PCR
Real-Time PCR analysis was undertaken using Roche Diagnostics' Light-Cycler 480 real-time PCR instrument and Light-Cycler 480 SYBR Green I Master kit, structured in 96-well plates. The SYBR Green I dye in the reaction mixture has specificity for double-stranded DNA. At each DNA synthesis stage, this dye binds to the amplified PCR products, which emit fluorescence upon binding, enabling detection of the amplicon.

To enhance PCR yield, sensitivity, and specificity, we utilized Hot Start protocols in conjunction with the Light-Cycler 480 SYBR Green I Master kit. These protocols employ the FastStart Taq DNA Polymerase, which has certain amino acid residues blocked with heat-labile groups. These groups inactivate the enzyme at ambient temperatures, thereby preventing nonspecific binding during the primer annealing phase. Pre-incubation at +95°C for 5 minutes activates the FastStart Taq DNA Polymerase, removing these groups and enabling DNA elongation during amplification. Each reaction was conducted in triplicate and included 5 μl SYBR Green I Master, 2 μl PCR-grade water, 2 μl cDNA, and 0.5 μl of each of the 10 μM forward and reverse gene-specific primers, resulting in a total volume of 10 μl. The reactions underwent a thermal cycling program involving an initial denaturation step at 95°C for 5 minutes, followed by 40 cycles of denaturation at 95°C for 15 seconds, annealing at 56.2°C for 15 seconds, and extension at 72°C for 20 seconds. We utilized the intercalating SYBR green assay as the reporter system in our study. The SYBR green intercalates between adjacent base pairs of double-stranded DNA and upon light excitation emits a fluorescent signal when bound to DNA. As the PCR cycles advance, the fluorescence signal intensifies, corresponding to the accumulation of amplicons. To ensure that the fluorescence signal only stemmed from the target templates and not from any nonspecific PCR product formation, we performed post-PCR dissociation curve analysis (melting curve analysis) ranging from 60 to 95°C. The fluorescence data was captured using the Light-Cycler 480 software (version 1.5; Roche Diagnostics). To enable advanced relative quantification across the genotypes and the three stages of stigma development, we employed the 2-ΔΔCt method proposed by Livak and Schmittgen in 2001. This method enables the comparison of gene expression levels based on Ct (cycle threshold) values and normalization against reference genes.

The amplified genes along with their forward and reverse primer sequences are displayed in Table 1. The CsTUB gene was amplified as an internal control, using the AMVRT cDNA kit from Roche Applied Science, Penzberg, Germany, as per the instructions in the user manual. For precision, the experiments were conducted twice. Following this, 5 μl of the PCR products were loaded onto a 1.2% (w/v) agarose gel (Sigma-Aldrich, St. Louis, MO, USA).

Table 1. Sequence and amplicon size of primers used for real time PCR analysis.

| Primer | Forward primer | Reverse primer | Amplicon size (bp) |
| --- | --- | --- | --- |
| CsZCD | GTCTTCCCCGACATCCAGATC | CTCTATCGGGCTCACGTTGG | 241 |
| CsLYC | AGATGGTCTTCATGGATTGGAG | ATCACACACCTCTCATCCTCTTC | 247 |
| CsBCH | TCGAGCT TCGGCATCACATC | GCAATACCAAACAGCGTGATC | 495 |
| CsGT2 | GATCTGCCGTGCGTTCGTAAC | GATGACAGAGTTCGGGGCCTTG | 400 |
| CsTUPB | TGATTTCCAACTCGACCAGTGTC | ATACTCATCACCCTCGTCACCATC | 225 |

## RESULTS AND DISCUSSION
### CsZCD expression during stigma development

Using RNA isolated from stigmas at distinct stages, this study employed reverse transcription and real-time PCR techniques. It predominantly delved into the CsZCD gene, investigating its mRNA fluctuation throughout the stigma's maturation process (refer to Figure 1). Notably, the zenith of the CsZCD gene expression was found during the scarlet phase, echoing the observations made by Castillo et al. (2005). Additionally, an examination of CsZCD's expression across each developmental point (as seen in Figure 1) suggests a tangible link with the concentration of apocarotenoids at those stages, hinting at a coordinated dance between the gene's expression and apocarotenoid build-up, a notion also proposed by Bustin (2000).

Expanding on this, real-time PCR was used to amplify both CsZCD and tubulin genes during the yellow, orange, and scarlet phases. The data revealed a 2.69-fold uptick in CsZCD gene activity compared to the tubulin gene in the scarlet phase. In contrast, the orange and yellow stages saw reductions of 0.90-fold and 0.69-fold, respectively, relative to tubulin. Moreover, the PCR data indicated an 8% rise in CsZCD activity from the yellow to orange stage and a noteworthy 33% jump from orange to scarlet. With the yellow phase expressing merely 25% of what was observed in the scarlet phase, there's a marked amplification in CsZCD activity transitioning from orange to scarlet. These observations mirror the findings from earlier reverse transcription PCR studies in saffron by Castillo et al. (2005) and Rubio et al. (2009).

However, this research distinguishes itself by introducing fold variations in CsZCD gene activity concerning an internal control during stigma evolution. Furthermore, it's the first to highlight a relationship between CsZCD activity and apocarotenoid aggregation. It's noteworthy that previous attempts to quantify the relative expression of CsZCD via real-time PCR in saffron are absent from literature. The implications of this study are profound, given its potential to steer commercial saffron spice production towards a desired apocarotenoid concentration. Grasping the gene dynamics of essential players in apocarotenoid synthesis during the development of stigmas is pivotal to maximize biotechnological avenues in boosting saffron yield. Moving forward, it's imperative to authenticate reference genes and delve into other significant genes like lycopene cyclase and β-carotene hydroxylase, setting the stage for congruent data in subsequent explorations.

### Mevalonate pathway (MVA) genes expression
RT-PCR was employed to investigate the semi-quantitative expression of CsZCD and CsLYC genes during the scarlet stage of stigma maturation, with CsTUB serving as the reference control. Notably, there were slight differences between the two genotypes; however, the CsZCD gene was more predominantly expressed than the CsLYC gene, as

# Gene Expression Patterns of CsZCD and Apocarotenoid Accumulation during Saffron Stigma Development

illustrated in Figure 2. Even though reverse transcription PCR primarily offers semi-quantitative data about gene expression, its importance cannot be understated in drawing comparative conclusions about gene expression levels. The method boasts impressive sensitivity and specificity, proving essential in pinpointing rare transcripts or in cases with limited sample availability.

Additionally, Quantitative real-time PCR (Q-PCR) techniques were also evaluated. This advanced method recognizes and measures target templates by observing the PCR product's growth, as evidenced by an accompanying fluorescence increase with every PCR cycle. This approach allows for accurate gene or transcript counts during the PCR amplification's exponential phase, correlating it directly with the preliminary counts of present target sequences. Unlike the traditional "end-point" PCR that only assesses amplicons after the PCR process is completed, measuring during the exponential phase mitigates potential issues (as discussed by Smith and Osborn, 2009). In terms of C. sativus, the developmental journey of stigma coincides with the switch from amyloplasts to chromoplasts, as well as the creation and storage of apocarotenoids. These are intrinsically connected to the expression trends of the CsZCD and CsLYC genes (as cited by Bouvier et al., 2003; Rubio-Moraga et al., 2009). This research scrutinized the accumulation trends of apocarotenoids, including key compounds like crocetin, picrocrocin, and various forms of crocin. This was done in mature saffron stigmas, noting variations in crocin concentrations and the lengths of stigmas, detailed in Table 2

**Table 2. Variability in stigma length of different saffron (Crocus sativus) selections.**

| Selections | Stigma length (cm) |
| --- | --- |
| CITH-S-125 | 3.74 ±0.04 |
| CITH-S-123 | 4.38 ±0.06 |
| CITH-S-124 | 3.86±0.05 |
| CITH-S-122 | 3.98±0.04 |
| CITH-S-12 | 3.44±0.05 |
| CITH-S-121 | 4.14±0.05 |
| CITH-S-107 | 4.84±0.02 |
| CITH-S-120 | 3.86±0.05 |
| CITH-S-104 | 3.72 ±0.04 |
| CITH-S-117 | 3.3±0.03 |
| CITH-S-112 | 3 ±0.06 |
| CITH-S-113 | 3.16±0.05 |
| CITH-S-119 | 2.98±0.04 |
| CITH-S-118 | 3.22±0.07 |
| CITH-S-10 | 2.9±0.05 |
| CITH-S-103 | 3.04±0.04 |
| CITH-S-43 | 3.16±0.05 |
| CITH-S-114 | 3.3±0.11 |
| CITH-S-115 | 3.2±0.05 |
| CITH-S-105 | 3.08±0.04 |
| CITH-S-106 | 3.34±0.09 |
| CITH-S-102 | 3.14±0.08 |
| CITH-S-108 | 3.4±0.03 |
| CITH-S-11 | 3.3h±0.03 |
| CITH-S-116 | 2.86 ±0.03 |
| CITH-S-13 | 3.42±0.05 |
| CITH-S-101 | 3.7 ±0.03 |
| CITH-S-3 | 3.3h±0.03 |
| CITH-S-111 | 3.12±0.07 |
| CITH-S-110 | 2.92±0.04 |
| CITH-S-76 | 3.2±0.03 |

The CsZCD gene plays a central role in orchestrating the synthesis of crocetin glucosides and picrocrocin, products that result from the action of zeaxanthin cleavage dioxygenase (Rubio-Moraga et al., 2009). Even though the build-up and makeup of carotenoids throughout stigma maturation are tightly regulated by the synchronized transcriptional activation of the genes related to carotenoid biosynthesis, there was an apparent discrepancy. The

                                                                            

## Gene Expression Patterns of CsZCD and Apocarotenoid Accumulation during Saffron Stigma Development

expression trends of the CsZCD and CsLYC genes did not mirror the storage patterns of apocarotenoid compounds. This indicates that the production of these molecules might be overseen by a different mechanism, possibly linked to the expression of carotenoid cleavage dioxygenase (referenced in Rubio-Moraga et al., 2008; Baghalian et al., 2010).

Recognizing the pivotal role of zeaxanthin cleavage in amassing apocarotenoids, the researchers chose to contrast the expression of the CsZCD gene in two separate saffron varieties. Additionally, this study also delved into the CsLYC gene, responsible for coding lycopene β-cyclase, the enzyme that transforms lycopene into β-carotene.

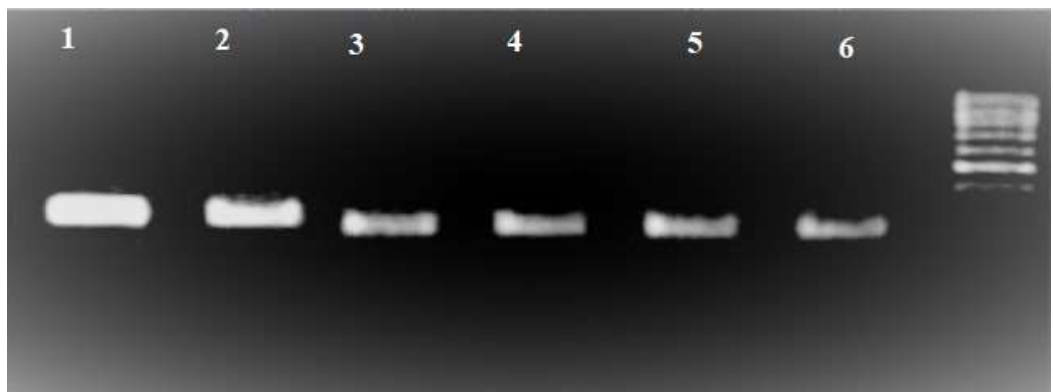

**Figure 1. At three different developmental stages of saffron stigmas, gene expression levels of *CsZCD* (Lane 4–6) and the reference gene Tubulin (Lane 1–3) were assessed.**

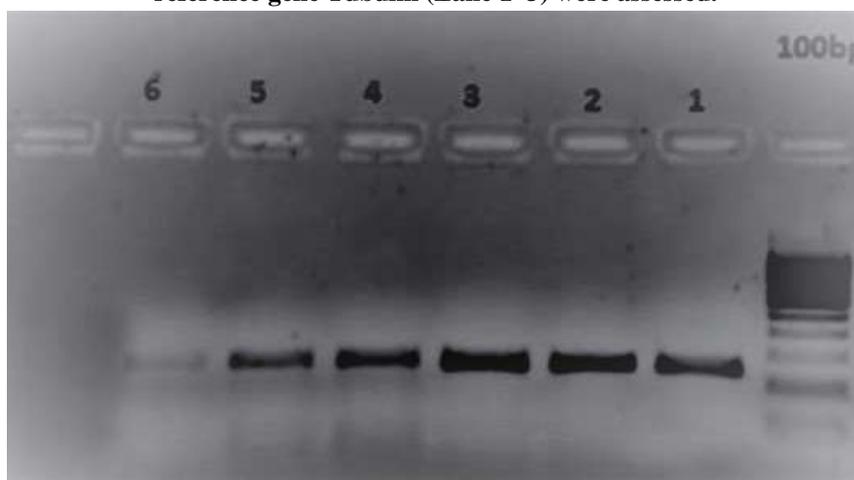

**Figure 2. Semi-quantitative analysis through reverse transcription PCR (RT-PCR) was performed to compare the gene expressions of CsTUB (Lane 1 and 2), CsLYC (Lane 3 and 4), and CsZCD (Lane 5 and 6) in the PAM-S-116 and CITH-S-107 saffron genotypes, respectively.**

In saffron's apocarotenoid production process, β-carotene and zeaxanthin act as pivotal forerunners through the mevalonate (MVA) route. During the scarlet phase of stigma development, real-time PCR amplified the CsZCD, CsLYC, and Tubulin genes for both genotypes (as illustrated in Figure 2). Remarkably, CsZCD displayed an expression rate that was three times greater than that of the CsTUB gene. On the other hand, CsLYC showed expression levels at 0.7-fold and 0.3-fold when compared to CsTUB.

These findings align well with past studies, especially the one conducted by Mir et al. in 2012. That study observed a 2.69-fold increase in CsZCD gene expression compared to the tubulin gene during the scarlet phase, alongside 0.90-fold and 0.69-fold decreases at the orange and yellow stages, respectively. The pronounced expression of the CsZCD gene, coupled with the simultaneous surge in apocarotenoid concentration as the stigma matures, highlights its influential role in both apocarotenoid formation and saffron stigma development (as referenced in Mir et al., 2012a, b). This data suggests that the difference in apocarotenoid content between the two genotypes could be attributed to variations in expression patterns of the CsZCD and CsLYC genes.

### CONCLUSION

This study aimed to delve into the expression dynamics of the CsZCD gene as saffron stigma matures, especially its linkage to apocarotenoid production. Throughout this developmental journey, the CsZCD gene exhibited significant expression variations, reaching its zenith during the scarlet phase, emphasizing its critical role therein. A strong correlation was found between the gene's expression and apocarotenoid build-up at each developmental milestone, further substantiating its key regulatory role. Real-time PCR provided a detailed quantitative picture, showcasing an 8%

# Gene Expression Patterns of CsZCD and Apocarotenoid Accumulation during Saffron Stigma Development

rise in CsZCD gene expression transitioning from yellow to orange, and a substantial 33% leap from orange to scarlet. Additionally, when juxtaposing CsZCD and CsLYC gene expressions, CsZCD emerged as more dominant, suggesting its plausible role in influencing apocarotenoid disparities. Such observations resonate with prior research, accentuating CsZCD's central role in apocarotenoid creation. In essence, this exploration unveils fresh insights into the controlling strategies of CsZCD throughout the saffron stigma's evolution and its ramifications on apocarotenoid deposition. Such revelations pave the way for potential advancements in commercial saffron spice production with precise apocarotenoid concentrations and set the stage for biotechnological innovations to boost saffron yields. However, to gain a holistic view of saffron stigma's growth and its cherished components, continued in-depth study is imperative, encompassing the validation of reference genes and the scrutiny of other significant genes.